# The (re-)instrumentalization of the *Diagnostic and Statistical Manual of Mental Disorders* (*DSM*) in psychological publications: a citation context analysis

Kai Li

School of Information Resource Management, Renmin University of China

islanderee@hotmail.com

**Abstract:**

Research instruments play significant roles in the construction of scientific knowledge, even though we have only acquired very limited knowledge about their lifecycles from quantitative studies. This paper aims to address this gap by quantitatively examining the citation contexts of an exemplary research instrument, the *Diagnostic and Statistical Manual of Mental Disorders* (*DSM*), in full-text psychological publications. We investigated the relationship between the citation contexts of the *DSM* and its status as a valid instrument being used and described by psychological researchers. We specifically focused on how this relationship has changed over the *DSM*'s citation histories, especially through the temporal framework of its versions. We found that a new version of the *DSM* is increasingly regarded as a valid instrument after its publication; this is reflected in various key citation contexts, such as the use of hedges, attention markers, and the verb profile in sentences where the *DSM* is cited. We call this process the *re-instrumentalization* of the *DSM* in the space of scientific publications. Our findings bridge an important gap between quantitative and qualitative science studies and shed light on an aspect of the social process of scientific instrument development that is not addressed by the current qualitative literature.

## 1 Introduction

Research instruments, i.e., *objects that are instrumental to scientific works*, are an important class of material objects involved in scientific research. Studies on scientific practices have found that instruments serve as a fundamental device through which researchers can gain access to "nature" that is otherwise invisible to human beings (Fraassen, 2008; Rheinberger, 1997) and as an important epistemological foundation of scientific objectivity (Daston & Galison, 2010). As quantitative researchers strive to understand the scientific system from broader perspectives (Leydesdorff, Ràfols, & Milojevic, 2020), the representation of scientific material objects, especially instruments, in scientific citations and texts has become an important topic yet to be fully investigated. Specifically, with an increasing number of publications addressing the scientific impact of research datasets and software entities (Howison & Bullard, 2015; Li, Yan, & Feng, 2017; Zhao, Yan, & Li, 2018), two key examples of scientific instruments, we need deeper knowledge about how the use and development of these research instruments are situated in epistemic cultures of science (Knorr-Cetina, 1999).

To address this gap, this paper presents an analysis of the citation contexts of a classic research instrument in the scholarship of mental disorder, the *Diagnostic and Statistical Manual of*

2*Mental Disorders* (*DSM*), as a response to the concept of *instrumentalization* in the field of science and technology studies (STS). Works by Bruno Latour and his colleagues (Latour, 1987; Latour & Woolgar, 1979) successfully established a materialist and practice-oriented tradition in science studies. In this line of research, a frequently recurring theme is that the status of research instruments is contingent and dependent on specific research contexts that are both temporal and local. Two major arguments have been proposed: that there are highly blurry and fluid boundaries between research instruments and other types of research objects (Engeström, 1990; Rheinberger, 1997) and that scientific knowledge and instruments are always co-produced (Jasanoff, 2004). These ideas are well summarized by Clarke and Fujimura (1992), who famously stated that scientific instruments are constructed through stabilization of scientific knowledge: one knowledge object becomes a tool when it is "no longer questioned, examined, or viewed as problematic, but is taken for granted" (pp. 10-11).

Given this conceptualization of instrumentalization, this study aims to examine how the *DSM* gained the status of a well-accepted research instrument from the perspective of scientific texts and citations. Developed by the American Psychiatric Association (APA), the *DSM* is one of the most widely used classification systems for mental disorders. Originally designed as a tool for inter-hospital communication in the 1950s, the *DSM* was gradually developed into a diagnostic scheme used by "psychiatrists, other physicians, and other mental health professionals that described the essential features of the full range of mental disorders" (American Psychiatric Association, 2013). Bowker and Star (2000) offered the observation that the *DSM* serves as the *lingua franca* for medical insurance companies, because of a lack of competing standards in the market. Moreover, the *DSM* is publicly regarded as an authoritative resource for both research and teaching in such fields as psychiatry and psychology, especially in North America (Millon & Klerman, 1986; Young, 1997).

According to Clarke and Fujimura (1992), temporality is an important scale along which research objects show different levels of instrumentality. This is the focus of the present study: we aim to understand how the level of instrumentality of the *DSM* has changed over its citation histories. The plural form of *history* is used here because there are multiple citation histories of the *DSM*: while we can take the *DSM* as a single object being cited in the scientific literature, each of its different versions may be deemed to have its own citation history.

**Table 1: Versions of the *DSM***

| Version | Abbreviation | Publication Year |
|---|---|---|
| 1st Edition | 1 | 1952 |
| 2nd Edition | 2 | 1968 |
| 3rd Edition | 3 | 1980 |
| 3rd Edition (Text Revision) | 3-TR | 1987 |
| 4th Edition | 4 | 1994 |
| 4th Edition (Text Revision) | 4-TR | 2000 |
| 5th Edition | 5 | 2013 |

In this study, we used various citation contexts as representations of the *DSM*'s instrumentality. First proposed in the early 1980s (Small, 1982), citation context analysis deals with the "particular message or statement within the citing document containing the reference" (p. 288), so that deeper meanings of citations can be extracted from publications. This method has frequently been adopted by researchers in quantitative science studies dating back to the 1980s, from Garfield's analysis of how Robert Merton's works are cited in different knowledge domains (Garfield, 1980) to McCain's Mean Utility Index, which considered both the location and context of citations (McCain & Turner, 1989), and her study of the longitudinal citation contexts of Frederick Brooks' book *The Mythical Man-Month* (McCain & Salvucci, 2006). The present research is particularly inspired by Small's recent work on the citation context of highly-cited method papers (Small, 2018), wherein a strong correlation was observed between the use of hedging phrases and particular verbs and whether or not the reference is cited in the method section.

In this research, we likewise assume that more frequent use of the *DSM* (or a specific *DSM* version) in the method section of research articles demonstrates a concomitantly higher level of instrumentality. Based on this assumption, we have extended the research framework adopted by Small (2018) and examined a broader set of linguistic attributes in citation sentences (citances) where the *DSM* is cited, from more than 100,000 full-text psychology research articles included in the Elsevier Text and Data Mining service. We hope our study is a first step towards a more profound appreciation of our scholarly communication system from a material-oriented perspective, which will help us shift away from document-centric bias and construct a fairer reward system for all kinds of scholarly outputs.

## 2 Method

### 2.1 Sample

To conduct this analysis, we acquired all English research articles in psychological journals included in the Elsevier Text and Data Mining (TDM) service[1], Elsevier's official API service offering access to all contents in the Elsevier ScienceDirect full-text database. This collection includes both open-access and institution-subscribed articles. Up to 2018, the ScienceDirect platform has the full text of over 15 million publications from more than 20,000 journals published as early as 1823[2]. This research is conducted on the full Elsevier TDM dataset as collected by the Centre for Science and Technology Studies (CWTS) at Leiden University. Using this database, the following data collection steps were undertaken.

First, all 203 journals classified under psychology were acquired on April 15, 2019. All English research articles published in these journals, as collected by CWTS, were included in our sample. A total of 203,803 unique DOIs were retrieved from this step.

---

[1] https://www.elsevier.com/about/policies/text-and-data-mining
[2] https://www.elsevier.com/__data/assets/pdf_file/0005/53528/0597-ScienceDirect-Factsheet-v4-HI-no-ticks.pdf

Second, from the above collection, we selected only the 145,892 papers with structured full-text data for inclusion in the final sample.

Third, we included only papers with at least one method section. The following four criteria were used to classify paper sections as either Method or non-Method:

- Criterion #1: A Method section title should contain the following terms: "experimental," "methodology," "methodologies," "method," or "methods;"
- Criterion #2 A Method section title must not contain the terms "experiment" or "experiments;"
- Criterion #3: Any subsection under a Method section is a Method section; and,
- Criterion #4: A Method section can be a subsection of a non-Method section.

This list of criteria considers the fact that psychological papers sometimes contain parallel sections with titles like "Experiment 1" and "Experiment 2." These sections serve as the container for a separate set of method, results, and sometimes discussion sections. We tested some additional keywords in our query, especially "data," but given that these failed to produce in any significant increase in the number of retrieved paper sections, we maintained our original query.

This approach to classifying paper sections was tested on the full paper sample and was evaluated manually. Its accuracy was judged to be satisfactory. After it was applied on our paper sample, I used all 22 method-section titles that (1) were manually validated and (2) appeared at least 100 times in the sample. These titles have in total 105,220 instances (in 104,094 unique papers); both figures are comparable to the 119,166 instances of the Introduction section found in our sample. All 104,094 articles were thus included in our final sample for this analysis. Based on this sample, we acquired all citances parsed by CWTS from their database.

Fourth, based on our final sample, we identified all references related to the *DSM*. In the CWTS database, a reference key is given to every unique indexed reference. This key is composed of (1) the name of the leading contributor, (2) the publication year, and (3) some other identifying information (such as the first few characters of the object title). For example, one key for the fifth edition of the *DSM* is "*americanpsychiatricassoci_2013_the*." However, non-publication objects are frequently cited with variant titles and other metadata elements (Li, Chen, & Yan, 2019), which explains why multiple keys (e.g. "*americanpsychiatricassoci_2013_ame*") were found for most *DSM* versions. To address this issue, I used the following criteria to integrate keys representing the same *DSM* version.

#1: The author of the reference is the American Psychiatric Association ("americanpsychiatricassoci");

#2: The object must be published in a year in which a major *DSM* version was published (as in Table 1);

#3: The initial letters of the title must be either "the," "dia," "dsm," or "ame," so that it is possible to cover different name forms of the *DSM* without including other resources published by APA in the same years. The selected letters cover such titles as "The Diagnostic and Statistical Manual of Mental Disorder" and "Diagnostic and Statistical Manual of Mental Disorder."

In total, 87 unique keys were identified for the seven major versions of the *DSM*. It should be noted that there are a few other keys that meet criteria #1 and #3, but not #2, i.e., keys with a different year from those in Table 1. These may be correct references that were given a wrong publication year mistakenly or intendedly. However, I decided not to use them because of the difficulty of accurately classifying them into the seven versions.

## 2.2 Measurements

Based on the sample discussed above, the following measurements were examined in this study.

### Ratio of citations in the method section

It has been commonly accepted that different sections within a paper have distinct narrative functions and significance (Swales, 1990). An assumption made in this study is that citations in the method section are more strongly connected to scientific instrumentality; this assumption is supported by various prior works in quantitative science studies (Bertin, Atanassova, Sugimoto, & Lariviere, 2016; Thelwall, 2019) and applied linguistics (Huang, 2014; Kanoksilapatham, 2012). By measuring the ratio of citations used in the Method section and how this ratio shifts over time, we strive to illustrate the extent to which a specific version of the *DSM* is regarded as an established research instrument, how this pattern changes over time, and how well this temporal trend is correlated with other linguistic attributes of *DSM* citances.

### Number of times a reference is cited in a paper

Voos & Dagaev (1976) reported that the number of citations to the same reference in a paper is indicative of the relationship between the citing and cited documents. There has been contradictory evidence since then. For example, Hooten (1991) found that multiple mentions of a reference are correlated with a closer association between the citing and cited documents, while Hanney and colleagues (2010) did not find any support for this conclusion.

### Linguistic attributes of citances

Linguistic attributes have been increasingly studied as a type of citation context (Ding, Liu, Guo, & Cronin, 2013; Jha, Jbara, Qazvinian, & Radev, 2017; Small, 2018). In this study, we focused on five classes of interactional markers identified by Hyland (1999), i.e., hedges, boosters, attitude markers, self-mentions, and engagement markers. They are, according to Hyland, resources available to writers to interact with their readers through their writings—for instance, in expressing their views or acknowledging uncertainties. Hedges are the only category that has been thoroughly examined in quantitative studies. As a result, the present study aims to offer a more comprehensive analysis of how these resources are used in the textual description of scientific instruments. The five resources are discussed below:

- **Hedges:** Hedges, such as *possible* and *perhaps*, are devices indicating uncertainties in writing, e.g., in acknowledging an alternative interpretation. Earlier quantitative studies have proven that hedges are inversely related to the method section and method-oriented references, because of the certain tone that is supposed to be expressed in this section (Chen, Song, & Heo, 2018; Small, 2018; Small, Boyack, & Klavans, 2019).
- **Boosters:** Contrary to hedges, boosters, such as *obviously* and *demonstrate*, are used to express certainties. Despite this opposition of function, Hyland (2005b) commented that the excessive use of either type of marker is discouraged by the research community.

- **Attitude markers:** Attitude markers are those words or phrases that express the writer's subjective attitudes toward the topics. Examples include attitude verbs (*agree*), sentence adverbs (*unfortunately*), and adjectives (*remarkable*). While no research has focused on their usage in scientific writings, we can assume that attitudes, like uncertainties, are less likely to be expressed in the method section than in other paper sections.

- **Self-mention:** Self-mention includes first-person pronouns (*we*) and possessive adjectives (*our*). In an earlier study, Hyland (2003) interpreted the ratios of self-mention in scientific publications across multiple research fields using the different research and persuasion strategies employed in these communities. For example, in the hard sciences, uniformities in the procedures and results are more important in convincing readers than is personal authority; thus, the texts in these fields are less personal in style than in the soft sciences. Despite such disciplinary differences, there is a lack of discussion as to how self-mention phrases are used differently between paper sections.

- **Engagement markers:** Engagement markers are devices to guide readers' attention or include them in the discourse. Such expressions include *by the way*, *you may notice*, and *note*.

**Readability of citances**

Readability is a frequently used linguistic attribute in evaluating the writing styles of scientific publications (Hartley, Pennebaker, & Fox, 2003; Hayden, 2008). Despite recent criticism (Hartley, 2016), we used the Flesch Scale in this study, as it is still widely used in quantitative studies of texts and especially scientific texts, as a marker for the differences between individual texts (Didegah, Bowman, & Holmberg, 2018; Oleinik, Kirdina-Chandler, Popova, & Shatalova, 2017; Van Wesel, Wyatt, & ten Haaf, 2014).

**Verbs used in citances**

Verbs bear strong rhetorical functions in the construction of scientific texts and are especially useful for understanding the relationship between citing and cited documents (Bertin et al., 2016; Bloch, 2010). Several studies have evaluated what verbs are the most frequently used in the method section (Bertin & Atanassova, 2014; Lamers, van Eck, Waltman, & Hoos, 2018; Small, 2018).

## 2.3 Analysis method

Using the 87 keys mentioned above, I acquired all citances of the *DSM* and conducted the following analyses.

For the linguistic analysis, I used the Spacy library in Python (Honnibal & Montani, 2017) to parse the sentences and extract verbs from them. Based on the statistics reported on its website, Spacy's parser reaches an accuracy of 94.48% as tested on a *Wall Street Journal* dataset.[3] Moreover, this library has been increasingly used to examine scientific corpora in recent years (Lamurias & Couto, 2019; Roth & Basov, 2020). In this study, I extracted and analyzed only the *primary verbs* in citances, or the verbs in the main clause. If there were multiple main verbs in the same sentence, I included all of them.

To determine whether a sentence used any of the five interactional resources identified by Hyland, I used the list of phrases offered in his book (Hyland, 2005a). All the included phrases were matched with the citances described above.

I used the Flesch Reading Ease Scale to calculate the readability of each citance. This domain-independent readability scale was developed by Rudolf Franz Flesch in 1943 (Flesch, 1948) and considers (1) sentence length and (2) the number of syllables to calculate the ease of reading for a corpus. It assigns a score from 0–100 to a corpus, with 100 representing the easiest and 0 the most difficult.

## 2.4 Description of the sample

In aggregate, the identified keys representing the *DSM* are cited in 17,695 citances belonging to a total of 12,435 papers. Figure 1 summarizes the ratio of citing articles among all sampled articles ($n = 145{,}892$) in a given year, and thus the relative importance of the *DSM* in our general paper sample. Despite the growing numbers of citations over time, the results show that the *DSM* has been cited in a relatively stable proportion of papers over the citation window.

**Figure 1: Ratio of articles citing the *DSM* over time**

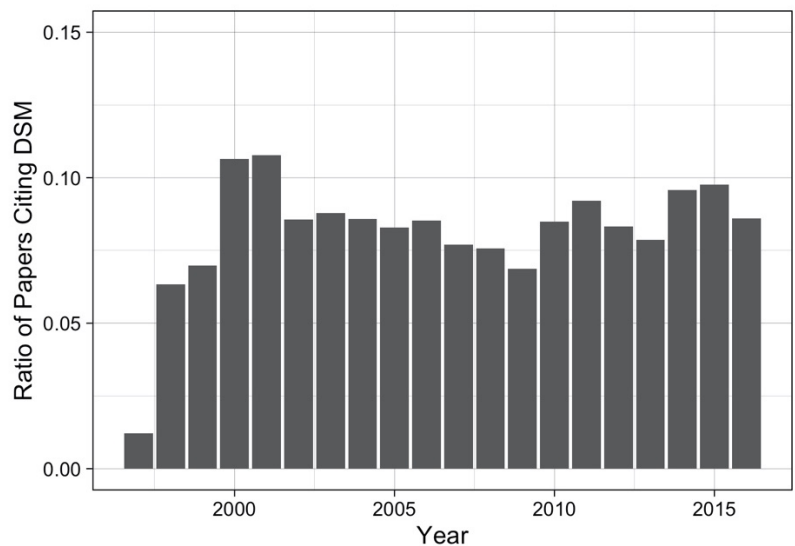

I also broke up the total number of citing publications by *DSM* version. The result is shown in Figure 2, with the y-axis representing the ratio of papers citing each version among the total number of papers in our overall sample. I used the object history (the year difference between a version's publication date and the citation date) as the x-axis, so that the patterns for all *DSM* versions can be standardized against their different relationships with the citation window.

---

[3] https://spacy.io/usage/facts-figures

**Figure 2: Ratios of citing articles citing specific versions of the *DSM***

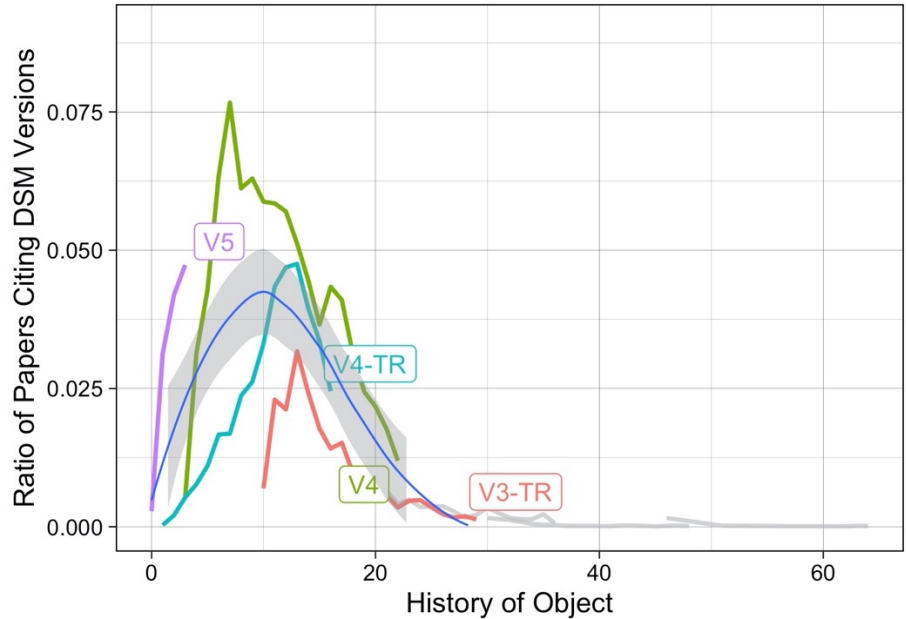

In our research design, only two *DSM* versions are fully covered by the citation window (from 1997 to 2016): Version 4-TR (2000) and Version 5 (2013). Moreover, Version 4 was published just ahead of the citation window (1994). We can see that each of these three versions is increasingly cited in its early years, despite the very different slopes. It is also obvious that, after a new version is published, fewer citations are given to the older versions. This can be explained by the fact that up-to-dateness is an important factor for researchers in selecting an instrument like the *DSM*.

A major focus of this study is to understand how linguistic attributes are used differently in citances over the histories of the *DSM*. Table 2 summarizes the measurements discussed above. In this table, as well as the rest of the study, I included only versions from V3-TR to V5 because of the small numbers of citations the first three *DSM* versions received (17, 29, and 399, respectively). The table shows that some measurements have strong variances among these four versions, which is the starting point of this study.

**Table 2: Summary of key attributes of *DSM* versions**

|  | **V3-TR** | **V4** | **V4-TR** | **V5** |
|---|---|---|---|---|
| No. of citing papers | 980 | 5419 | 3890 | 1709 |
| No. of citing papers with method-section citance | 64.8% | 63.7% | 53.2% | 16.8% |
| No. of citances | 1143 | 7630 | 5725 | 2702 |
| No. of citances in method section | 59.8% | 52.1% | 42.1% | 11.8% |
| Citances per paper | 1.17 | 1.4 | 1.47 | 1.58 |
| Ratio of citances with attention markers | 2.3% | 4.4% | 5.6% | 7.5% |
| Ratio of citances with boosters | 9.4% | 8.7% | 9.2% | 8.9% |
| Ratio of citances with self-mentions | 3.4% | 4.6% | 3.6% | 5.2% |
| Ratio of citances with engagement markers | 21.7% | 20.6% | 19.9% | 20.3% |
| Ratio of citances with hedging | 10.7% | 16.2% | 19.6% | 26.6% |
| Mean readability score | 22.62 | 18.96 | 15.56 | 13.86 |

## 3 Results

### 3.1 How often is the *DSM* cited in the method section?

One of the most notable differences among *DSM* versions in Table 2 is the ratio of citances used in the method section. The likelihood for a citation to be given in this section decreased significantly from Version 3-TR to Version 5, as shown in Figure 3, where each data point is an aggregated ratio for a specific year. (This figure omits data points where a version has fewer than 10 citances in a year, to reduce radical outliers.)

**Figure 3: Ratio of method section citances by *DSM* version**

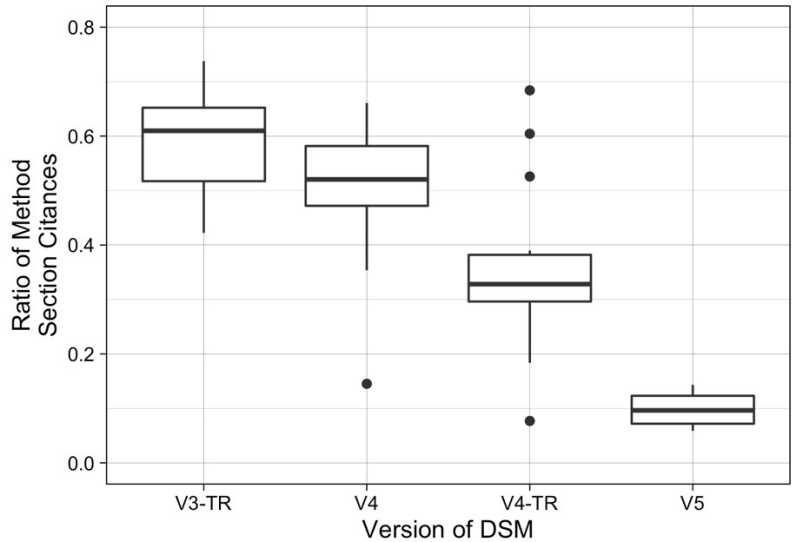

Given the fact that these four versions have distinct relationships with the citation window, I further evaluated whether the differences were caused by differences in the portion of the lifecycle covered by the citation window. In Figure 4, I plotted how the method-section citance ratio changed over each version's citation history. There seems to be a converging trend among the four versions, where the ratio increases over the early years of each object, then gradually decreases around the 15th year, based on versions 3-TR and 4. This result is strong indication that the *DSM*, as a well-developed scientific instrument, still takes time to be regarded as an instrument that can be reliably used in scientific research in the space of scientific publications.

**Figure 4: Ratio of method section citances over years by *DSM* version**

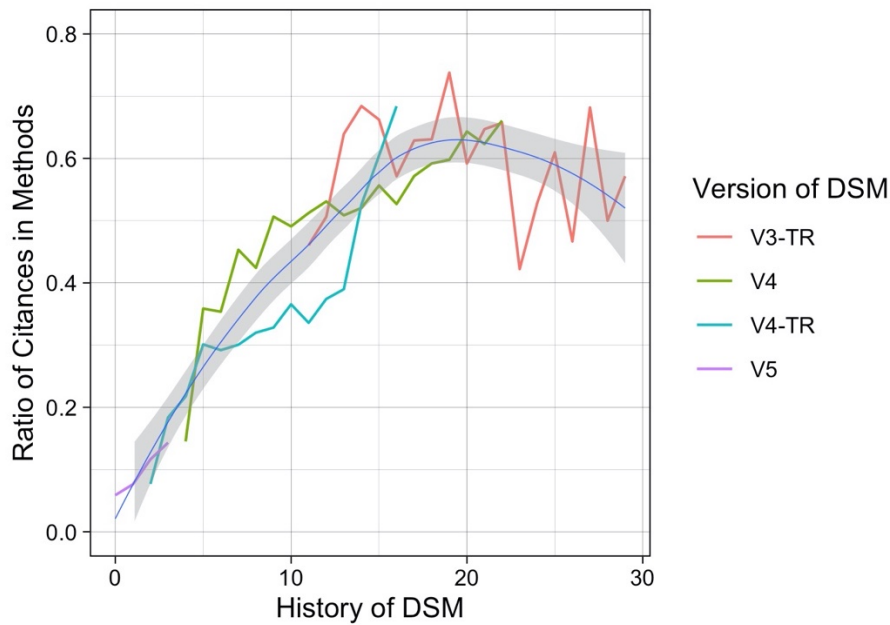

**3.2 Density of citance in papers**

Another pattern with a strong linear trend among various versions is the number of citances per paper. As shown in Figure 5, from versions 3-TR to 5, the mean number of citances per paper keeps increasing, while that in the method section decreases significantly.

**Figure 5: Mean number of citances per paper by *DSM* version**

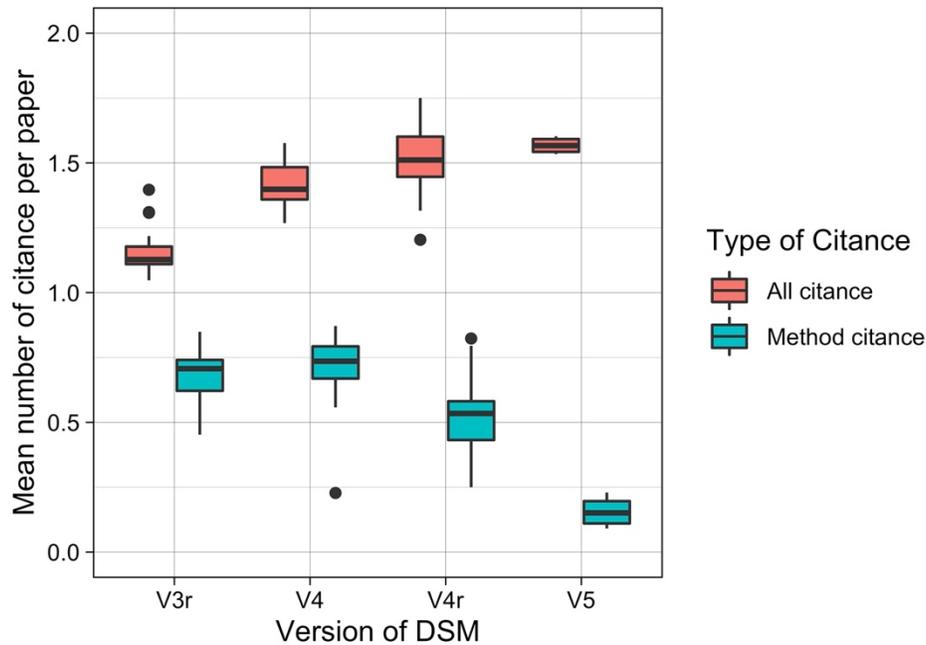

Figure 6 shows how both trends change over the citation histories of all *DSM* versions. They are once again correlated with the different lifecycles of these versions. In its beginning years, a new version is mentioned in non-method sections more frequently but barely mentioned in the method section. In light of the results from the previous section, the decreasing number of citances in non-method sections in the earlier years may be explained by the need for researchers to introduce or justify the *DSM* as a research instrument after it is published. This need, of course, gradually diminishes as the *DSM* is increasingly accepted as an established instrument.

**Figure 6: Mean number of citances per paper by *DSM* history**

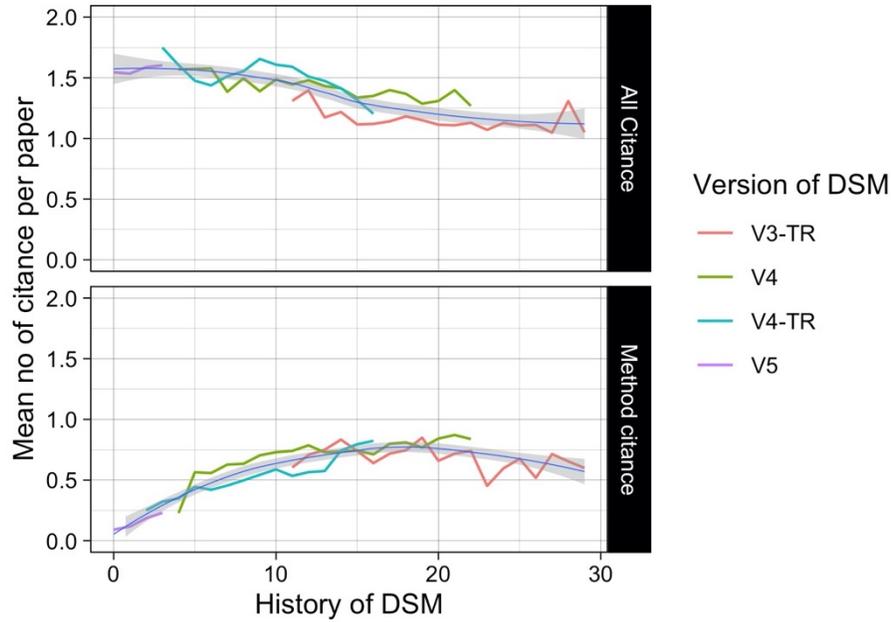

## 3.3 Linguistic analysis

We further examined the linguistic attributes related to *DSM* citances. Figure 7 illustrates how the five interactional markers are used differently over the citation histories of *DSM* versions.

**Figure 7: Mean ratio of citances with Hyland's interactional markers**

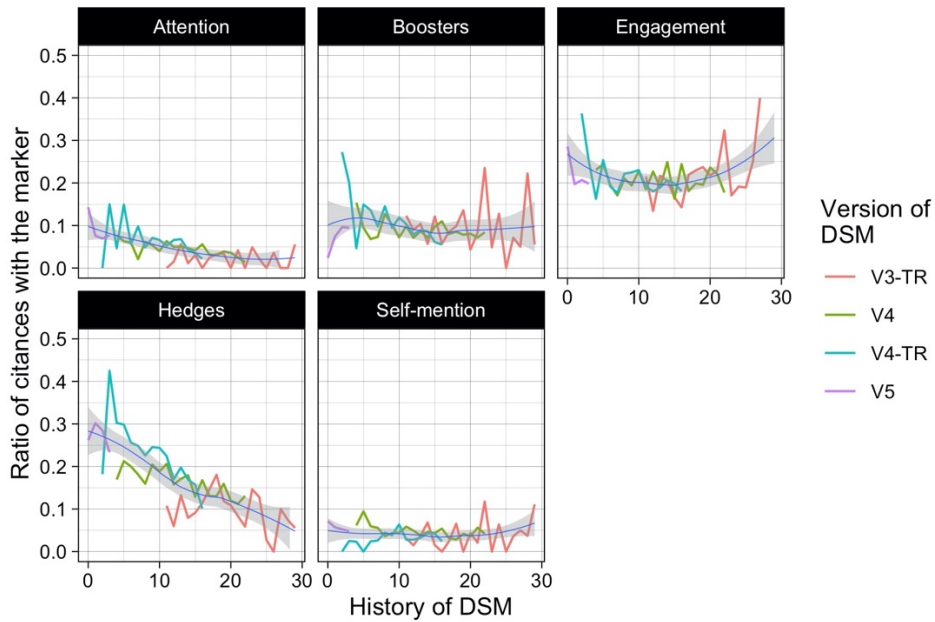

As shown in Figure 7, hedges and attention markers are the only two categories with linear changes over the citation history. The decreasing trend in both variables is consistent with our overall assumption that, as the *DSM* is more established, it is increasingly used as an instrument and thus is described with a lower level of uncertainty (decreasing use of hedges) and in more factual tones (decreasing use of attention markers). The trends for self-mention markers and boosters are essentially flat, and that for engagement markers appears to be subject to radical fluctuation, even though we removed all data points with fewer than 10 citances. The same can be said for the readability score as shown in Figure 8 below.

**Figure 8: Mean readability score by *DSM* history**

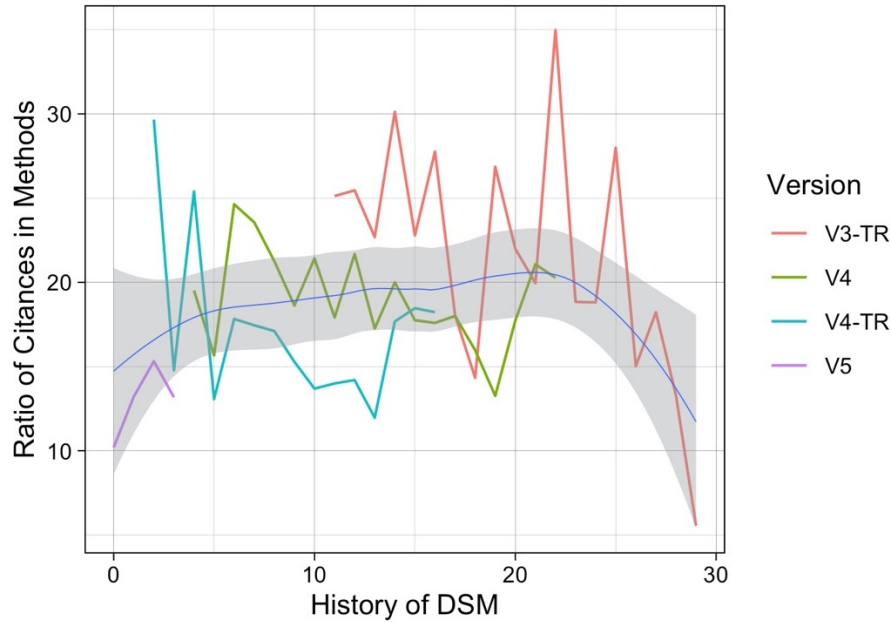

On the other side of the story, some variables examined above display quite different usage patterns in method and non-method sections, respectively, even though they may or may not show any temporal trends. The results are summarized in Table 3, where the "All versions" category contains only versions from V3-TR onward, instead of all seven versions of DSM.

**Table 3: Summary of linguistic attributes by *DSM* version**

| Version | Section | Attention | Boosters | Self-mention | Engagement | Hedging | Readability |
|---|---|---|---|---|---|---|---|
| V3-TR | Method | 1.3% | 7.4% | 3.6% | 21.4% | 7% | 22.77 |
| V3-TR | Non-Method | 4.3% | 13.4% | 3% | 22.2% | 18.3% | 22.3 |
| V4 | Method | 1.4% | 6% | 4.3% | 20% | 6.9% | 22.83 |
| V4 | Non-Method | 8.6% | 12.5% | 4.9% | 21.6% | 29.4% | 13.39 |
| V4-TR | Method | 2% | 6.6% | 4% | 20.8% | 7.6% | 22.11 |
| V4-TR | Non-Method | 8.8% | 11.5% | 3.2% | 19.1% | 30.2% | 9.78 |
| V5 | Method | 1.3% | 7.5% | 9.7% | 27.3% | 15.1% | 28.45 |
| V5 | Non-Method | 8.5% | 9.2% | 4.5% | 19.2% | 28.4% | 11.65 |
| All version | Method | 1.5% | 6.4% | 4.5% | 20.8% | 7.5% | 22.84 |
| All versions | Non-Method | 8.4% | 11.3% | 4% | 20.1% | 28.8% | 12.25 |

A few observations can be drawn from this table. First, both hedges and attention markers are used differently across paper sections and over time. Given the fact that there is a higher ratio of citations being given in the method section as a reference grows older, this result is consistent with our findings in Figure 7. On average, a citance in a non-method section is almost three times more likely to have a hedging phrase and four times more likely to have an attention marker than one in the method section, which makes these features strong predictors of the citation function of a reference. Moreover, there is an especially high use of hedges in citances of the *DSM-V* in the method section (15.1%), showing that even though it is in the method section, there is still a higher level of uncertainty being expressed in the very early years of this version.

Second, like hedging and attention markers, boosters and the readability score exhibit different patterns in the method section than in other paper sections. Specifically, boosters are also used more frequently in non-method sections; this is consistent with Hyland's comment that even in the method section, the expression of certainty is discouraged, even though it is not consistent with our findings regarding the temporal trend.

### 3.4 Verbs used in citances

The 10 most frequently used verbs extracted from all citances, method section citances, and non-method section citances are summarized in Table 4.

**Table 4: Top 10 verbs from our sample**

| Rank | All | Method | Non-Method |
|---|---|---|---|
| 1 | include | include | characterize |
| 2 | meet | meet | include |
| 3 | characterize | diagnose | define |
| 4 | diagnose | assess | classify |
| 5 | use | use | consider |
| 6 | assess | make | report |
| 7 | define | base | suggest |

| 8  | make   | recruit   | use      |
|----|--------|-----------|----------|
| 9  | base   | consist   | diagnose |
| 10 | recruit| establish | find     |

It can be observed that our lists of verbs are very different from the top verbs reported in other similar works (Bertin & Atanassova, 2014; Small, Tseng, & Patek, 2017), where verbs that are most frequently used in the method section normally include *use*, *perform*, *follow*, etc. Most of these verbs rank very low in our results. Moreover, many verbs in our lists do not fall into the category of *research verbs* (i.e., verbs aiming to describe the procedure or acts of research). However, the top verbs identified from this study are consistent with the distinct ways in which the *DSM* is involved in research: it is a standard used to diagnose a mental disorder and to support decisions about recruiting participants.

Notwithstanding the differences between our verb lists and those of previous works, we found that verbs could function as a valid means of measuring the text similarity of citances between the method and non-method sections. We calculated the frequencies of all 153 verbs that appeared in our sample at least 10 times (with "be" and "have" removed because they are used too broadly). To measure the similarity between these two sections, we applied Spearman rank correlation to the ranks of verbs from the two corpora. Table 5 summarizes the relationships between the three broad citance groups: those in both the method and non-method sections, those in the method section only, and those in the non-method sections only. The result shows that the rankings of verbs in the method and non-method sections are strongly negatively correlated with each other.

**Table 5: Spearman rank correlation among verb lists from method and non-method citances**

| Category | Rho value |
|---|---|
| All citances - Method citances | 0.708 |
| All citances - Non-Method citances | 0.509 |
| Method citances - Non-method citances | –0.547 |

Moreover, I also compared how verbs from each version of the *DSM* are distributed relative to the above-mentioned verb lists. The results, summarized in Table 6, show that verb distributions in newer *DSM* versions are more similar to those in non-method sections and vice versa. This strongly supports our earlier observation that *DSM* versions are increasingly used in the method section by showing that the use of verbs can be reliably used to evaluate the instrumentality of references, even though the use of individual verbs may not work very well because of the low density of these verbs in citances.

**Table 6: Spearman rank correlation between verbs in specific versions and general lists**

| Category | All to Method | All to Non-Method | Method to Method | Non-Method to Non-Method |
|---|---|---|---|---|
| V3-TR | 0.707 | 0.175 | 0.674 | 0.502 |
| V4 | 0.799 | 0.298 | 0.865 | 0.91 |
| V4-TR | 0.602 | 0.507 | 0.774 | 0.892 |
| V5 | 0.030 | 0.801 | 0.546 | 0.877 |

## 4 Discussion

### 4.1 The re-instrumentalization of the *DSM* in scientific texts

The present research offers a case study of how citation contexts shift over the citation histories of the *DSM* in the field of psychology. By conducting a citation context analysis, we examined the relationship between the instrumentality of the *DSM* and various linguistic attributes, such as the use of interactional markers and verbs in the citation sentences.

The most interesting finding from this research is the fact that, even though every version of the *DSM* was meticulously developed into a blackboxed instrument (Regier, Narrow, Kuhl, & Kupfer, 2009), it does not automatically become one after being published, at least in the space of scientific publications. Instead, it still takes time for the *DSM* to be accepted as a valid instrument by researchers. This is primarily shown by the fact that as a new *DSM* version matures, it is increasingly used in the method section compared to its earlier years. This general conclusion is supported by existing works on the rhetorical functions of paper sections (Bertin et al., 2016; Huang, 2014; Kanoksilapatham, 2012) and is also consistent with the citation contexts of sentences in different paper sections, such as the increasing use of *DSM* citations in the method section and the opposite trend for the use of hedges and attention markers in *DSM* citances. The fact that these patterns are shown for all major *DSM* versions adds further validity to our conclusion.

This conclusion sheds light on the processes of the construction of scientific instruments in two significant ways. First, it aims to bridge the gap between quantitative and qualitative science studies on the topic of scientific instruments. This gap was largely created by various barriers to proper and sufficient representations of research instruments in citation data (Li et al., 2019, 2017). As a result, quantitative researchers have only gathered very limited evidence about the performance of these material objects in the scholarly communication system, even less so from the perspective of their lifecycles. This reality makes it very difficult for a material-oriented perspective to be established in quantitative science studies and thus for more conversations to transpire between these two research communities. Second, our evidence shows that, after its physical development and publication, the *DSM* does not automatically become a research instrument. This process of re-instrumentalization is not covered by STS literature concerning how research instruments are produced, which shows how quantitative evidence can help to expand theories developed in qualitative communities.

Moreover, in our discussion of the process of re-instrumentalization, we have also shown the significant roles played by the temporal framework in the appreciation of the citation context—in particular, how multiple time frames may exist in an object's citation history. The concept of version is a vital yet highly under-examined topic in quantitative science studies; versions are critical to the identification of many non-publication objects, such as research datasets (Pröll & Rauber, 2013) and software objects (Smith, Katz, & Niemeyer, 2016). Moreover, versions are gaining in relevance as pre-publication paper repositories are increasingly used by researchers from nearly every field, which creates multiple versions of research articles (Larivière et al., 2014). This study offers some preliminary evidence about the roles played by versioning in the citation history of an object. Specifically, we have shown that each version of the *DSM* can be treated as a unique epistemic object, with its own citation history and similar patterns of citation contexts. However, to more deeply understand the concept of version in scholarly communication, we plan to conduct more studies in the future to elucidate how it may work differently for other types of research object and in different knowledge domains.

**4.2 Citation contexts of research instruments**

Another major contribution of the present work is to offer a more comprehensive examination of the relationships between citation contexts and method-related rhetorical functions, inspired by recent works concerning citation contexts of method papers[4] (Small, 2018; Small et al., 2019). Our results supported Small's key finding that hedges are a central predictor for method-related citation context, especially the distinctively low level of uncertainties expressed in the method section. However, apart from hedges, we also evaluated how other interactional phrases are used along the lifecycle of the *DSM*. We found that attention markers are another strong indicator for how the *DSM* is cited. Like hedges, these show strong variances not only between the method and non-method sections, but also between different stages of the *DSM*'s citation histories. In addition to hedges and attention markers, boosters are shown to be used differently between the method section and other paper sections. However, this difference does not translate into temporal patterns. Instead, the use of boosters is relatively stable over the citation histories of the *DSM*.

The different patterns for these interactional markers point to the complexity of human language in scientific writings. One example of such complexity is that citations to the *DSM* are not consistently given in an instrumental context. This may be applicable to other research instruments, given the diversity of citation practices for material objects (Li et al., 2019). To address this complexity, these linguistic attributes can be employed as useful instruments for future works to automatically identify research instruments from scientific citations and texts, which will further help to construct a material-oriented history of science.

This study also evaluated the distribution of main verbs in all citances of the *DSM*. Two conclusions are drawn from the analysis. First, the verb profile of the *DSM* (even when based on all method section citances) is very different from earlier findings based on all citations in the method section (Bertin et al., 2016). This finding indicates that even under the citation contexts of the method section, there are different and highly individual connections between individual citations (or instruments) and verbs (or actions taken in the research). This is consistent with how the method section is conceptualized in the Create A Research Space (CARS) model, where the section serves to describe a localized research setting (Swales, 1990).

---

[4] Many, if not most, of these method papers are representations of research instruments, especially when they are cited in the method section, even though "research instrument" is not the framework adopted by Dr. Henry Small.

This individualized connection between verbs and citations raises questions about the extent to which we can use a general verb ranking or one scheme of citation context to analyze the method section, given the vast diversity of research actions taken by researchers. Before fully transforming verbs into a valid research instrument, we need to build better knowledge about the categorization of research actions expressed by action verbs, with a comprehensive consideration of local factors, such as disciplinarity and the lifecycle of the research objects represented by the citation. This research direction will be an important complement to existing works on full-text scientific publications.

Despite the individual nature of the connections between verbs and citations, verbs have proven to be a solid instrument to evaluate citation contexts at a somewhat aggregated level. By analyzing the similarity of verb profiles, I draw the same conclusion that older *DSM* versions are more strongly connected to the method section than are newer ones.

## 5 Conclusion

The research reported in this paper offers a citation context analysis of how the *DSM* is cited in full-text psychological publications. Our results show that the instrumentality, or the extent to which the *DSM* is used in the method section, varies as specific versions of the *DSM* go through their respective life stages. Over the first few years after its publication, a *DSM* version is increasingly cited in the method section, which indicates it is increasingly cited as a research instrument over time. This ratio does not seem to be influenced by the publication of the next version and only reaches its peak around 10–15 years. Moreover, this changing level of instrumentality is accompanied by shifts in the use of some other important citation contexts within citation sentences. We found significantly different uses of hedges, attention markers, and some verbs as the *DSM* becomes more mature.

Our study offers an important quantitative examination of how citations serve as research instruments, one citation function that has not been extensively studied in quantitative science studies. Specifically, this research build connections between quantitative evidence and theories of instrumentalization developed in qualitative studies in the field of STS. We found another aspect of how research instruments are constructed that is not fully addressed in the qualitative literature: after "physical" packaging, research instruments need to be re-instrumentalized in the space of scientific texts, as reflected in those texts and their citations.

This paper, despite the significance of its findings, represents only the first step towards a deeper understanding of how research instruments are represented in scientific texts. After all, it is only a case study of a unique research instrument on many different levels. To address this limitation, as the next step of this research project, we plan to conduct larger-scale quantitative analyses to more comprehensively examine the findings from the present study, especially the manner in which these research instruments are used differently across knowledge domains.

22
headings. *Journal of Informetrics*, *13*(2), 555–563.

Van Wesel, M., Wyatt, S., & ten Haaf, J. (2014). What a difference a colon makes: how superficial factors influence subsequent citation. *Scientometrics*, *98*(3), 1601–1615.

Voos, H., & Dagaev, K. S. (1976). Are All Citations Equal? Or, Did We Op. Cit. Your Idem?. *Journal of Academic Librarianship*, *1*(6), 19–21.

Young, A. (1997). *The harmony of illusions: Inventing post-traumatic stress disorder* (Vol. 11). Princeton University Press.

Zhao, M., Yan, E., & Li, K. (2018). Data set mentions and citations: A content analysis of full-text publications. *Journal of the Association for Information Science and Technology*, *69*(1), 32–46. https://doi.org/10.1002/asi.23919